\documentclass{PoS}
\usepackage[utf8]{inputenc}

\usepackage{amsmath,amssymb}
\usepackage{graphicx}
\usepackage{epsfig}
\usepackage{multirow}
\usepackage{float}
\usepackage{booktabs}

\def\proof{\noindent{\sl Proof:}\kern0.6em}

\def\frac#1#2{\hbox{$#1\over#2$}}
\def\dual{\mathstrut^*\kern-0.1em}

\def\lvec#1{\setbox0=\hbox{$#1$}
    \setbox1=\hbox{$\scriptstyle\leftarrow$}
    #1\kern-\wd0\smash{
    \raise\ht0\hbox{$\raise1pt\hbox{$\scriptstyle\leftarrow$}$}}
    \kern-\wd1\kern\wd0}
\def\rvec#1{\setbox0=\hbox{$#1$}
    \setbox1=\hbox{$\scriptstyle\rightarrow$}
    #1\kern-\wd0\smash{
    \raise\ht0\hbox{$\raise1pt\hbox{$\scriptstyle\rightarrow$}$}}
    \kern-\wd1\kern\wd0}
\def\slash#1{\setbox0=\hbox{$#1$}\setbox1=\hbox{$\kern1pt/$}
    #1\kern-\wd0\kern1pt/\kern-\wd1\kern\wd0}

\def\nabstar#1{{\nabla\kern0.5pt\smash{\raise 4.5pt\hbox{$\ast$}}
               \kern-5.5pt_{#1}}}

\def\nabbarstar#1{{\overset{\leftarrow}{\nabla}\kern0.5pt\smash{\raise 4.5pt\hbox{$\ast$}}
               \kern-5.5pt_{#1}}}

\def\drvstar#1{{\partial\kern0.5pt\smash{\raise 4.5pt\hbox{$\ast$}}
               \kern-6.0pt_{#1}}}

\def\ldrvstar#1{{\lvec{\,\partial}\kern-0.5pt\smash{\raise 4.5pt\hbox{$\ast$}}
               \kern-5.0pt_{#1}}}

\def\MeV{{\rm MeV}}
\def\GeV{{\rm GeV}}

\def\MSbar{\overline{\rm MS\kern-0.5pt}\kern0.5pt}
\def\Nf{{N_{\rm f}}}

\def\psibar{\overline{\psi}}

\def\zetabar{\bar{\zeta}}
\def\zetaprime{\zeta\kern1pt'}
\def\zetabarprime{\zetabar\kern1pt'}

\def\diracstar#1#2{
    \setbox0=\hbox{$\gamma$}\setbox1=\hbox{$\gamma_{#1}$}
    \gamma_{#1}\kern-\wd1\kern\wd0
    \smash{\raise4.5pt\hbox{$\scriptstyle#2$}}}

\def\Ds{D_{\rm s}}
\def\DsdagDs{\Ds{\Ds}^{\kern-1pt\dagger}}

\def\avg#1{{\kern1.0pt\overline{\kern-1.0pt#1\kern-1.0pt}\kern1.0pt}}

\title{Towards a precise determination of the equation of state 
	   of QCD at high-temperature}

\ShortTitle{Towards the EoS of QCD at high-temperatures}

\author{\speaker{Mattia Dalla Brida} \\
        Dipartimento di Fisica, Universit\`a di Milano-Bicocca,
        and INFN, Sezione di Milano-Bicocca, 
        Piazza della  Scienza 3, I-20126 Milano, Italy \\
        E-mail: \email{mattia.dallabrida@unimib.it}}

\author{Leonardo Giusti \\
        Dipartimento di Fisica, Universit\`a di Milano-Bicocca,
        and INFN, Sezione di Milano-Bicocca, 
        Piazza della  Scienza 3, I-20126 Milano, Italy \\
        E-mail: \email{Leonardo.Giusti@cern.ch}}

\author{Michele Pepe \\
        INFN, Sezione di Milano-Bicocca, 
        Piazza della  Scienza 3, I-20126 Milano, Italy \\
        E-mail: \email{Michele.Pepe@mib.infn.it}}

\abstract{
  We present preliminary results towards a fully non-perturbative determination 
  of the equation of state of QCD at very high-temperatures, $T\approx 3-80\,\GeV$. 
  The key ingredient is the lattice formulation of QCD in a moving reference frame, 
  which allows us for a neat determination of the entropy density from simple 
  expectation values of the momentum-components of the energy-momentum tensor. 
  For the computation we employ $\Nf=3$ flavours of non-perturbatively O($a$)-improved 
  Wilson-fermions. We present an analysis of the O($a$)-improvement of the expectation 
  values entering the determination, and show how these can be accurately computed in 
  simulations.
}

\FullConference{XIII Quark Confinement and the Hadron Spectrum - Confinement2018\\
		31 July - 6 August 2018\\
		Maynooth University, Ireland}

\begin{document}

\section{Introduction}

The equation of state (EoS) is a fundamental quantity of QCD as it describes the 
thermodynamic properties of quarks and gluons at equilibrium. It is relevant for a 
wide spectrum of applications, including the physics of the early universe, the study 
of neutron stars, and the description of heavy-ion collisions. Tremendous 
progress has been made over the past decade in the determination of the EoS at zero 
chemical potential for temperatures $T\approx 500\,\MeV$ and below, using lattice field 
theory methods~\cite{Borsanyi:2013bia,Bali:2014kia,Bazavov:2014pvz}. Lattice QCD is 
indeed the only known framework that allows for a first-principles, non-perturbative 
computation of the EoS. On the other hand, all current state of the art determinations
rely on (some variant of) the integral method~\cite{Boyd:1996bx}. This approach 
is very challenging from the computational point of view as one needs to accommodate
within the same simulation the physics at zero-temperature \emph{and} at the  
temperatures of interest. Due to this limitation, only very recently some first 
exploratory study touched temperatures  $T\approx1-2\,\GeV$~\cite{Bazavov:2017dsy}. 

In the past few years a new framework has been proposed for the determination of the 
EoS using lattice QCD. The idea is to consider QCD in a moving reference 
frame~\cite{Giusti:2010bb,Giusti:2011kt,Giusti:2012yj}.%
\footnote{For other interesting approaches we recommend the reader to 
		  refs.~\cite{Kanaya:2017cpp,Caselle:2018kap}.}
Within this framework one is able to completely decouple simulations at different
temperatures, which makes computationally feasible to reach high-temperatures 
while having all uncertainties under control. The method has been very successfully 
applied to the SU(3) Yang-Mills theory~\cite{Robaina:2013zmb,Giusti:2014ila,Umeda:2014ula,
Giusti:2015daa,Giusti:2016iqr}. The study of refs.~\cite{Giusti:2014ila,Giusti:2015daa,
Giusti:2016iqr}, in particular, obtained a determination of the EoS over two orders of 
magnitude in the temperature with half a per-cent accuracy. Following these encouraging 
results in the pure gauge sector, our goal is to obtain a similar determination in full QCD. 

A first exploratory study in this direction was presented in~\cite{DallaBrida:2017sxr},
where some first numerical experience with the method was described. In this
contribution we present some preliminary results towards a systematic determination 
of the EoS of $\Nf=3$ QCD in the completely unexplored regime of temperatures: 
$T\approx 3 - 80\,\GeV$. This determination is of great interest for several reasons. 
First of all, it fills the gap in our knowledge of the EoS between temperatures 
of the order of the typical QCD scales, up to the electro-weak scale. This information
is very relevant for understanding the physics of the early universe. Secondly, the 
results will provide an important, solid test of perturbation theory in the high-temperature 
regime. In this range of temperatures  perturbation theory is commonly used to approximate 
the EoS. On the other hand, the results in the SU(3) Yang-Mills theory of~\cite{Giusti:2014ila,
Giusti:2015daa,Giusti:2016iqr}, clearly point to the fact that this might not be as accurate 
as it is typically assumed. It is thus compelling to assess this issue in the most relevant
case of QCD. Finally, our computation intends to be the first determination of the EoS 
which systematically employs the Wilson rather than the staggered formulation of lattice QCD. 
Proving that precise results using Wilson-quarks are feasible with today's computational 
resources opens the way to further studies using this formulation. This is particularly relevant
in view of consolidating the current state of the art determinations at lower temperatures.
  
The outline of this contribution is the following. In the next section we briefly review the 
framework of lattice QCD in a moving frame, and recall how the EoS can be computed in terms
of simple expectation values of the energy-momentum tensor (EMT). In Sect.~\ref{sec:EMT}, 
we discuss our lattice set-up and give an analysis of the O($a$)-improvement 
of the relevant expectation values of EMT. In Sect.~\ref{sec:Results}, we present the 
lines of constant physics we plan to follow for our high-temperature determination 
of the EoS, and discuss some preliminary results for the bare expectation values of the EMT. 
We finally conclude with an outlook on future work.

\section{Thermodynamics from a moving frame} 
\label{sec:Thermodynamics}

Considering a thermal quantum field theory in a moving frame brings a new perspective 
in studying thermodynamics. This framework provides us indeed with new relations 
to compute thermodynamic potentials~\cite{Giusti:2010bb,Giusti:2011kt,Giusti:2012yj}. 
To define a thermal quantum field theory in a moving frame one must 
impose \emph{shifted boundary conditions} (SBC) on the fields integrated over in the 
Euclidean path-integral. In the case of lattice QCD this amounts to requiring: 
\begin{equation}
	\label{eq:SBC}
	U_\mu(L_0,\vec{x})=U_\mu(0,{\vec{x}}-{L_0}{\vec{\xi}}),
	\quad
	\psi(L_0,\vec{x})=-\psi(0,{\vec{x}}-{L_0}{\vec{\xi}}),
	\quad
	\psibar(L_0,\vec{x})=-\psibar(0,{\vec{x}}-{L_0}{\vec{\xi}}).
\end{equation}
In these equations, $L_0$ denotes the physical extent of the lattice in the  
temporal direction, while ${\vec{\xi}}$ is the so-called \emph{shift} vector 
corresponding to the Euclidean velocity of the moving frame; note that 
in the rest frame, $\vec{\xi}=0$, the usual thermal/periodic boundary 
conditions are recovered. We then consider periodic boundary conditions for all
fields in the three spatial dimensions of extent $L$.  

The entropy density of the system, $s(T)$, is the central quantity of interest.  
Once $s(T)$ is known, the other thermodynamic potentials including the pressure 
$p(T)$ and the energy density $\varepsilon(T)$ can be inferred from standard 
thermodynamic identities. In a moving reference frame the entropy density is 
related to the momentum density of the system by~\cite{Giusti:2012yj}:
\begin{equation}
	\label{eq:Entropy}
	{s(T)\over T^3}=-{L_0^4(1+{\vec{\xi}^2})^3\over \xi_k}\,
	\langle T^R_{0k} \rangle_\xi,
	\qquad
	T^{-1}=L_0\sqrt{1+\vec{\xi}^2},
\end{equation}
where in this equation $\langle \cdot\rangle_\xi$ denotes the lattice path-integral 
expectation value in the presence of SBC, and $T_{0k}^R$ is the momentum 
$k$-component of the renormalized EMT. Eq.~(\ref{eq:Entropy}) is  
the natural transcription to a quantum field theory of the corresponding classical 
relativistic relation (cf.~ref.~\cite{Landau:1982dva}). In the following we 
discuss some of its relevant features when studied on the lattice.

\section{The energy-momentum tensor on the lattice}
\label{sec:EMT}

\subsection{General considerations}

On the lattice the EMT requires renormalization due to the explicit breaking
of Poincar\'e symmetry by the regularization. The problem of defining 
a properly renormalized EMT which respects, up to discretization errors, the 
correct Ward identities, was first addressed in a series of pioneering 
papers~\cite{Caracciolo:1989pt,Caracciolo:1991vc,Caracciolo:1991cp}. Focusing 
on the momentum-components of the EMT, the analysis of these references shows 
that properly renormalized fields can be defined as:
\begin{equation}
	\label{eq:T0kR}
	T_{0k}^R(x)=Z_{F}(g_0)T^F_{0k}(x) +  Z_{G}(g_0)\, T^G_{0k}(x),
\end{equation}
where $T^F_{0k}$ and  $T^G_{0k}$ are the bare fermionic and gluonic components
of the EMT, respectively. The renormalization constants $Z_{F}$ and $Z_{G}$ 
appearing in this equation are renormalization scale independent, and thus only 
functions of the bare gauge coupling $g_0$. They depend on the specific discretization 
of the QCD action and of the bare fields $T^F_{0k}$ and $T^G_{0k}$,
and they are fixed by imposing the validity of some continuum Ward identity 
at finite lattice spacing; only in this way we are guaranteed that in the continuum 
limit $T_{0k}^R$ will converge to the generator of spatial translations. 
For our study we employ (non-perturbatively) O($a$)-improved Wilson-fermions and 
consider the Wilson gauge action~\cite{Yamada:2004ja}. For the bare EMT we take 
the discretized form given in eqs.~(4) and (5) of ref.~\cite{DallaBrida:2017sxr}. 

We have recently completed a 1-loop computation of the renormalization constants 
$Z_F$ and $Z_G$ within the framework of SBC. This calculation provides us with useful insight 
on how to choose practical renormalization conditions to fix $Z_{F}$ and $Z_{G}$ also 
non-perturbatively. In addition, the results give us  valuable perturbative information on 
discretization effects: both to determine improvement coefficients (cf.~Sect.~\ref{subsec:Improvement}), 
as well as to perturbatively subtract lattice artefacts from our non-perturbative determinations
of the $Z$-factors. The details of this computation will be presented elsewhere~\cite{Draft:2019}.
Earlier computations of some of these renormalization factors using different set-ups can be 
found in refs.~\cite{Caracciolo:1989pt,Caracciolo:1991vc,Caracciolo:1991cp,Capitani:1994qn,Burgio:1996ji}.

\subsection{O($a$)-improvement}
\label{subsec:Improvement}
 
Wilson-fermions are known to be affected by discretization errors of O($a$). If one aims at
precise continuum extrapolations it is hence convenient to systematically remove them following
Symanzik improvement programme~\cite{Symanzik:1983dc,Symanzik:1983gh}. In short, this consists 
in adding irrelevant O($a$)-counterterms with the proper symmetry transformations to the lattice 
action and fields, and to tune their coefficients to cancel the unwanted O($a$)-effects in 
on-shell quantities. Given the fact that our action is O($a$)-improved, improved expectation 
values of the EMT are obtained by improving the EMT itself. 

In this section we focus on the specific case of improving the expectation value $\langle T^{R}_{0k}\rangle_\xi$;
a more general discussion on the improvement of the EMT will be presented in~\cite{Draft:2019}. We then 
begin by considering the situation where the quarks are massless. The O($a$)-counterterms we need to 
add to $T^R_{0k}$ are given in principle by all mass dimension 5 operators which share the same lattice 
symmetries of these fields. Symmetry transformations and the field equations of motion can however be used 
to reduce the actual number of fields we need to consider in a specific (on-shell) expectation 
value~\cite{Luscher:1996sc}. Following the discussion of ref.~\cite{Capitani:2000xi}, it is possible to 
show that an O($a$)-improved definition of $\langle T^{R}_{0k}\rangle_\xi$ can be obtained as:
\begin{equation}
	\label{eq:T0kIMassless}
	\langle T^{R}_{I,0k}\rangle_\xi =
	 Z_G(g_0)\langle {T}^{G}_{0k}\rangle_\xi
	+Z_F(g_0)\big\{\langle {T}^{F}_{0k}\rangle_\xi+a\langle\delta {T}^{F}_{0k}\rangle_\xi\big\},  
\end{equation}
where for the fields $\delta {T}^{F}_{0k}$ we take:
\begin{equation}
	\label{eq:OaConterterm}
	\delta {T}^{F}_{0k}(x)
	=c_T^{F}(g_0)\frac{1}{8}\psibar(x)\big[\sigma_{0\rho}\widehat{F}_{k\rho}(x)+
	\sigma_{k\rho}\widehat{F}_{0\rho}(x)\big]\psi(x).
\end{equation}
In this equation, $\widehat{F}_{\mu\nu}$ denotes the (traceless) clover discretization of the 
field strength tensor (see e.g.~ref.~\cite{Giusti:2015daa} for its definition), while 
$c_T^{F}(g_0)$ is an improvement coefficient which must be tuned to remove the O($a$)-effects 
from eq.~(\ref{eq:T0kIMassless}).
 
When considering the case of massive quarks more O($a$)-counterterms need to be added to
the EMT. Here we discuss only the case of mass-degenerate quarks and leave the details of the 
non-degenerate case to~\cite{Draft:2019}. We also assume that the bare gauge coupling $g_0$ 
is properly improved, and so the subtracted bare quark-masses, $m_{\rm q}=m_0-m_{\rm cr}(g_0)$, 
with $m_0$ the bare quark mass and $m_{\rm cr}$ its critical value (see ref.~\cite{Luscher:1996sc} 
for a discussion about this point). In this situation, an O($a$)-improved definition of 
$\langle T^{R}_{0k}\rangle_\xi$ in the presence of mass-degenerate quarks is given by:
\begin{equation}
	\label{eq:TImunu}
	\langle T^{R}_{I,0k}\rangle_\xi =
	 Z_G(\widetilde{g}_0)\langle {T}^{G}_{I,0k}\rangle_\xi
	+Z_F(\widetilde{g}_0)\langle {T}^{F}_{I,0k}\rangle_\xi,  
\end{equation}
where $\widetilde{g}_0$ is the improved bare coupling~\cite{Luscher:1996sc},
and 
\begin{equation}
	{T}^{G}_{I,0k}(x) = 
	\big(1+b_T^{G}(g_0)am_{\rm q}\big){T}^{G}_{0k}(x),
	\quad
	T^{F}_{I,0k}(x) = 
	\big(1+b_T^{F}(g_0)am_{\rm q}\big)\big\{T^{F}_{0k}(x) + a\delta {T}^{F}_{0k}(x)\big\}.
\end{equation}
By a proper tuning of the $b$-coefficients, one can eliminate all 
O($am_{\rm q}$) discretization errors stemming from the EMT. It is easy to show that at tree-level 
in perturbation theory this is achieved by setting $b_T^G=0$ and $b_T^F=1$~\cite{Draft:2019}. 

The above discussion shows that, in general, in order to improve the expectation value 
$\langle T^{R}_{I,0k}\rangle_\xi$ we need to determine the improvement coefficients: 
$c_T^F$, $b_T^F$, and $b_T^G$. However, in the high-temperature regime of $\Nf=3$ QCD,
none of these coefficients is expected to be very relevant. At temperatures $T\gg 100\,\MeV$,
chiral symmetry is not broken for vanishing quark-masses; the only sources of chiral symmetry 
breaking at these temperatures are the quark-masses themselves. At high-temperature 
the chirally non-invariant O($a$)-counterterm (\ref{eq:OaConterterm}) hence contributes to 
$\langle T^{R}_{I,0k}\rangle_\xi$ with O($am_{\rm q}$)-effects; all discretization errors in 
$\langle T^{R}_{I,0k}\rangle_\xi$ are thus of O($am_{\rm q}$). Now, in order to keep discretization
effects under control we must always be in the situation where $aT\ll1$, which implies that:
$am_{\rm q}\ll m_{\rm q}/T$. Consequently, in $\Nf=3$ QCD, where only the up, down, and strange 
quarks are considered, in the range of temperatures of interest, $T\approx 3-80\,\GeV$, the 
${\rm O}(am_{\rm q})$-effects affecting $\langle T^{R}_{I,0k}\rangle_\xi$ are expected to be
well-below the per-cent level. In addition, small values of the lattice spacing, $a\ll T^{-1}$, 
correspond to relatively small values for the bare coupling $g_0$. Perturbative estimates for the 
improvement coefficients $c_T^F$, $b_T^F$, and $b_T^G$, are thus expected to be good enough
to have these small O($am_{\rm q}$) effects under control. We plan to corroborate all these 
expectations by studying the size of the relevant O($a$)-counterterms non-perturbatively,
so to be able to quantitatively estimate their effect in our results.

\section{Simulation strategy and results}
\label{sec:Results}

In this section we give some details on the lines of constant physics (LCPs) that
we plan to follow for the determination of the entropy density in the high-temperature 
range, $T\approx 3-80\,\GeV$. The aim is to obtain a final precision on the \emph{continuum} 
entropy density of about $1\%$ over the whole temperature range. The first step is to 
find values of the bare coupling $g_0$ and of the temporal resolution $L_0/a$ which correspond 
to a good set of constant physical temperatures to cover the desired range. To this end, 
we consider the LCPs defined by the ALPHA collaboration in terms of running couplings~\cite{Brida:2016flw,
DallaBrida:2016kgh,Bruno:2017gxd,DallaBrida:2018rfy,Campos:2018ahf} (see ref.~\cite{DallaBrida:2018cmc} 
for a recent review). Using in particular the results of refs.~\cite{DallaBrida:2018rfy,
Campos:2018ahf} we are able to fix the bare parameters for 8 values of the temperature, 
where for each of these we can have up to 4 lattice spacings, corresponding to the resolutions: 
$L_0/a=6,8,10,12$. As for the choice of shift vector we take $\vec{\xi}=(1,0,0)$. Perturbative 
studies suggest that for these values of $L_0/a$, this should result in small discretization 
errors for the entropy (cf.~ref.~\cite{DallaBrida:2017sxr}). Given this choice and the results 
of ref.~\cite{Bruno:2017gxd}, we can infer from the values of the renormalized couplings used 
to fix the temperatures the physical value of the latter. Table \ref{tab:Results} 
collects these results together with the bare parameters of the ensembles that we are currently 
generating. To conclude with the lattice geometry, we are considering for all ensembles a spatial 
resolution of $L/a=288$. This translates into having $TL\approx 34-17$ as we go from $L_0/a=6-12$. 
Since finite volume effects are exponentially small in $TL$~\cite{Giusti:2012yj}, we expect these 
effects to be well-below our target precision. We are currently conducting a systematic 
study to confirm that indeed finite volume effects in the relevant matrix elements of the EMT
are negligible within our statistical precision.

\begin{table}[hptb!]
  \centering
  \small
  \begin{tabular}{llllll}
  \toprule
  $T$ (GeV) & $L_0/a$ & $\beta$  & $\langle T_{0k}^G \rangle_\xi/T^4$ & 
  $\langle T_{0k}^F \rangle_\xi/T^4$ & $N_{\rm ms}$ \\
  \midrule
  $2.8$   & $6$  & $6.2735$ & $-2.361(19)$ & $-5.8667(94)$ & $100$ \\
  $2.8$   & $8$  & $6.4680$ & $-2.421(33)$ & $-5.700(18) $ & $250$ \\
  $4.6$   & $6$  & $6.6050$ & $-2.462(19)$ & $-5.9477(80)$ & $100$ \\
  $7.5$   & $6$  & $6.9433$ & $-2.562(18)$ & $-6.0265(95)$ & $100$ \\
  $11.9$  & $6$  & $7.2618$ & \multicolumn{2}{c}{in production} & $20$ \\
  $19.2$  & $6$  & $7.5909$ & $-2.650(25)$ & $-6.1757(85)$ & $100$ \\
  $30.5$  & $6$  & $7.9091$ & \multicolumn{2}{c}{in production} & $25$ \\
  $47.8$  & $6$  & $8.2170$ & $-2.707(18)$ & $-6.264(11) $ & $100$ \\
  $76.5$  & $6$  & $8.5403$ & $-2.806(23)$ & $-6.305(10) $ & $100$ \\
  $76.5$  & $8$  & $8.7325$ & $-2.784(40)$ & $-6.132(13) $ & $250$ \\
  \bottomrule
  \end{tabular}
  \caption{Values of the temperatures we are considering and corresponding 
		   values of $\beta=6/g^2_0$ and $L_0/a$ of the ensembles that we are 
		   currently generating. All simulations have $L/a=288$. Our preliminary 
		   results for the bare gluonic and fermionic matrix elements of 
		   the EMT are also given, together with the total number of 
		   (independent) measurements we collected. We do not observe any 
		   autocorrelation between measurements on successive trajectories
		   of length 2 MDUs.}
   \label{tab:Results}
\end{table}

For our choice of lattice action refs.~\cite{DallaBrida:2018rfy,Campos:2018ahf} give results for 
the critical value of the quark-masses, $m_{\rm cr}(g_0)$. We employ these values and conveniently
fix the bare quark masses $m_0=m_{\rm cr}(g_0)$ in our simulations. Up to discretization errors 
our 3-flavours of quarks can therefore be considered massless. Given the fact that physical quark-mass
effects in the entropy are of O($m_{\rm q}^2/T^2$), we expect that in our range of temperatures the difference 
with having physical values for the quark-masses is well-below our target accuracy. However, we plan 
to carefully study this issue by investigating the mass dependence of our results; this also in relation
to estimate the size of the O($am_{\rm q}$) errors one would have for physical values of $m_{\rm q}$. 

In table \ref{tab:Results} we listed our preliminary results for the bare matrix elements, 
$\langle T^{G}_{0k}\rangle_\xi$, $\langle T^{F}_{0k}\rangle_\xi$, for the ensembles we are
currently considering. Given the large lattice volume we are simulating we obtain very precise 
results with only a few hundred measurements. At fixed $L_0/a$ and number of  measurements the 
precision we reach does not seem to depend too strongly on the temperature. For $L_0/a=6$, we 
typically reach a precision on the gluonic matrix element of $\approx 0.6-0.9\%$, while for the 
fermionic contribution we have $\approx 0.15\%$ accuracy at all temperatures. When changing $L_0/a$ at 
fixed $\beta$, we (roughly) observe the expected scaling of the relative error with $(L_0/a)^4$. Finally, 
we may gain some rough idea on cutoff effects in the entropy density at the highest temperatures 
by taking for the renormalization factors $Z_F,Z_G$ their 1-loop values~\cite{Draft:2019}. 
If we consider in particular the ratio between the entropy density obtained for $L_0/a=6$ and $8$ at 
$T_{\rm pt}=76.5\,\GeV$, we find: $[s(T_{\rm pt},L_0/a=8)/s(T_{\rm pt},L_0/a=6)]|_{Z_{F,G}^{\rm 1-loop}}\approx 0.98$, 
which means that discretization errors are of the order of $2\%$. Bearing in mind that the non-perturbative 
$Z$-factors will also carry their cutoff effects and statistical errors, safe continuum limit 
extrapolations with $1\%$ accuracy seem at hand.

\section{Conclusions and outlook}

In this contribution we presented some preliminary results towards a per-cent accuracy 
determination of the EoS of $\Nf=3$ QCD in the range of temperatures: $T\approx 3-80\,\GeV$. 
The results we obtain for the bare matrix elements of the EMT are very encouraging: we can 
reach very high-precision with modest computational resources. In addition, the set-up allows 
us to have systematic effects under control, including both finite-volume and discretization 
effects. In order to better understand the nature and size of the latter we have studied in 
some detail the O($a$)-improvement of the relevant expectation values of the EMT. The next 
mandatory step at this point is the determination of the renormalization constants $Z_F$, $Z_G$. 
We already devised a set of non-perturbative renormalization conditions and tested their viability 
to 1-loop order in lattice perturbation theory~\cite{Draft:2019}. Their non-perturbative 
determination is on going.
 
\section{Acknowledgements}

The code used for the simulations is based on the \texttt{openQCD-1.6} package~\cite{Luscher:2012av,
LuscherWeb:2016}. All simulations were performed on the HPC cluster Wilson at the University of 
Milano-Bicocca, and on the Marconi machine at CINECA, through agreements of INFN and the University
of Milano-Bicocca with CINECA. We gratefully acknowledge the computer resources and the technical 
support provided by these institutions.

\bibliographystyle{JHEP}
\bibliography{confxiii} 

\end{document}